\documentclass[11pt]{article}  
\usepackage{times}
\usepackage{graphicx}

\usepackage{cite}
\usepackage{url}
\usepackage{mathfont}
\def\min{\mathop{\rm min}}
\def\max{\mathop{\rm max}}
\def\log{\mathop{{\rm log}}}
\let\union\cup

\mathcode`O="724F

\DeclareSymbolFont{AMSb}{U}{msb}{m}{n}
\DeclareSymbolFontAlphabet{\Bbb}{AMSb}
\def\R{\ensuremath{\Bbb R}}

\def\E{\ensuremath{\Bbb E}}
\def\H{\ensuremath{\Bbb H}}
\def\S{\ensuremath{\Bbb S}}
\DeclareMathSymbol{\subsetneq}{\mathrel}{AMSb}{"28}

\newtheorem{theorem}{Theorem}
\newtheorem{lemma}{Lemma}
\newtheorem{open}{Open Problem}

\DeclareSymbolFont{lasy}{U}{lasy}{m}{n}
\let\Box\undefined
\DeclareMathSymbol\Box{0}{lasy}{"32}
\newcommand{\qed}{\hfill$\Box$\medbreak}
\newenvironment{proof}{\noindent{\bf Proof: }}{\qed}

\begin{document}

\title{Optimal M\"obius Transformations\\
for Information Visualization and Meshing}
\author{Marshall Bern\thanks{Xerox PARC, 3333 Coyote Hill Rd., Palo 
Alto, CA 94304}\and
David Eppstein\thanks{Univ. of California, Irvine, Dept. Inf. \&
Comp. Sci., Irvine, CA 92697. Work done in part while visiting Xerox
PARC and supported in part by NSF grant CCR-9912338.}}
\date{ }
\maketitle

\begin{abstract}
We give linear-time quasiconvex programming algorithms for finding
a M\"obius transformation of a set of spheres in a unit ball
or on the surface of a unit sphere that maximizes the minimum size of a
transformed sphere.  We can also use similar methods to maximize the
minimum distance among a set of pairs of input points.  We apply
these results to vertex separation and symmetry display in spherical graph
drawing, viewpoint selection in hyperbolic browsing, element size
control in conformal structured mesh generation, and brain flat mapping.
\end{abstract}

\section{Introduction}

M\"obius transformations of $d$-dimensional space form one of the
fundamental geometric groups.  Generated by inversions of spheres, they
preserve spherical shape as well as the angles between pairs of curves
or surfaces.  Thus, M\"obius transformations provide a mechanism for
transforming a geometric input in a way that preserves some important
structure within the input while allowing other structure such as
object size to vary. We consider here problems of finding an {\em optimal
M\"obius transformation}, one that optimizes some objective function of
the transformed input.  Specifically, we consider the following problems:

\begin{itemize}
\item Given a set of $(d-1)$-dimensional spheres in Euclidean space
$\E^d$, all contained within a unit ball, find a M\"obius transformation
that maps the unit ball to itself and maximizes the minimum radius among
the transformed spheres.

\item Given a set of $(d-1)$-dimensional spheres on a sphere $\S^d$,
find a M\"obius transformation of $\S^d$ that  maximizes the minimum
radius among the transformed spheres.

\item Given a graph connecting a set of vertices on $\S^d$ or in the
unit ball in $\E^d$, find a M\"obius transformation that maximizes the
minimum length of an edge of the transformed graph.
\end{itemize}

We develop efficient algorithms for solving these problems, by
formulating them in terms of quasiconvex programming in a
hyperbolic space.  The same formulation also shows that a simple
hill-climbing approach, to search for a transformation that cannot be
improved by any small perturbation, is guaranteed to find the global
optimum; we do not analyze the time complexity of this approach but it
is likely to work well in practice as an alternative to our more
complicated quasiconvex programming algorithms.

We apply these results to the following areas:

\begin{itemize}
\item Spherical graph drawing~\cite{Koe-BSAW-36}.  Any embedded planar
graph can be represented as a collection of tangent circles on a sphere
$\S^2$; this representation is unique for maximal planar graphs, up to
M\"obius transformation.  By applying our algorithms, we can find a
canonical spherical realization of any planar graph that optimizes either
the minimum circle radius or the minimum separation between two vertices,
and that realizes any symmetries implicit in the given embedding.

\item Hyperbolic browsing~\cite{LamRaoPir-CHI-95}.  The Poincar\'e model
of the hyperbolic plane has become popular as a way of displaying web
sites and other graph models too complex to view in their entirety.  This
type of model permits parts of the site structure to be viewed in detail,
while reducing the size of peripheral parts.  Our algorithms
provide a way of finding a ``central'' initial viewpoint for a
hyperbolic browser, that allows all parts of the sites to be viewed at
an optimal level of detail.

\item Mesh generation~\cite{BerPla-HCG-00,ThoWarMas-85}.  A principled
method of structured mesh generation involves conformal mapping of the
problem domain to a simple standardized shape such as a disk,
construction of a uniform mesh in the disk, and then inverting the
conformal mapping to produce mesh elements in the original domain. 
M\"obius transformations can be viewed as a special class of conformal
maps that take the disk to itself.  By applying our optimal
transformation methods, we can find a conformal mesh that meets given
requirements of element size in different portions of the input domain,
while using a minimal number of elements in the overall mesh.

\item Brain flat mapping~\cite{HurBowSte-TR-99}.  Hurdal et al. have
proposed a system for visualizing convoluted brain surfaces, by finding
an approximate conformal mapping of those surfaces to a Euclidean disk,
sphere, or hyperbolic plane.  Our methods can be used to
choose a conformal mapping that minimizes the resulting areal distortion.
\end{itemize}

This paper is structured as follows.  We begin by briefly describing the
group of M\"obius transformations and outlining the results
from hyperbolic geometry and quasiconvex programming needed to describe
our algorithms (Section~\ref{sec:prelim}).  We describe
our algorithms for finding optimal M\"obius transformations
under various optimization criteria
(Sections \ref{sec:algs}), and finally
describe in more detail the applications outlined above
(Sections~\ref{sec:apps}).

For purposes of asymptotic analysis,
we assume throughout that the dimension $d$ of the spaces we deal with
is a constant; most commonly in our applications, $d=2$ or $d=3$.

\section{Preliminaries}
\label{sec:prelim}

\subsection{M\"obius Transformation and Hyperbolic Geometry}
\label{sec:mob}

An {\em inversion} of the set $\R^d\cup\{\infty\}$, generated by a
sphere $C$ with radius $r$, maps to itself every ray that originates
at the sphere's center. Within each
such ray, each point is mapped to another point along the same ray, so
that the product of the distances from the center to the point and to its
image equals
$r^2$.  The center of $C$ is mapped to $\infty$ and vice versa.
An inversion maps each point of $C$ to itself, transforms spheres to
other spheres, and preserves angles between pairs of curves or surfaces.
Repeating an inversion produces the identity mapping on
$\R^d\cup\{\infty\}$.

The set of products of inversions forms a group, the group of
{\em M\"obius transformations} on the Euclidean space $\E^d$.
If we restrict our attention to the subgroup that maps a given sphere
$\S^{d-1}$ to itself, we find the group of M\"obius transformations on
$\S^{d-1}$.

\begin{figure}[t]
\centering
\includegraphics[width=2.5in]{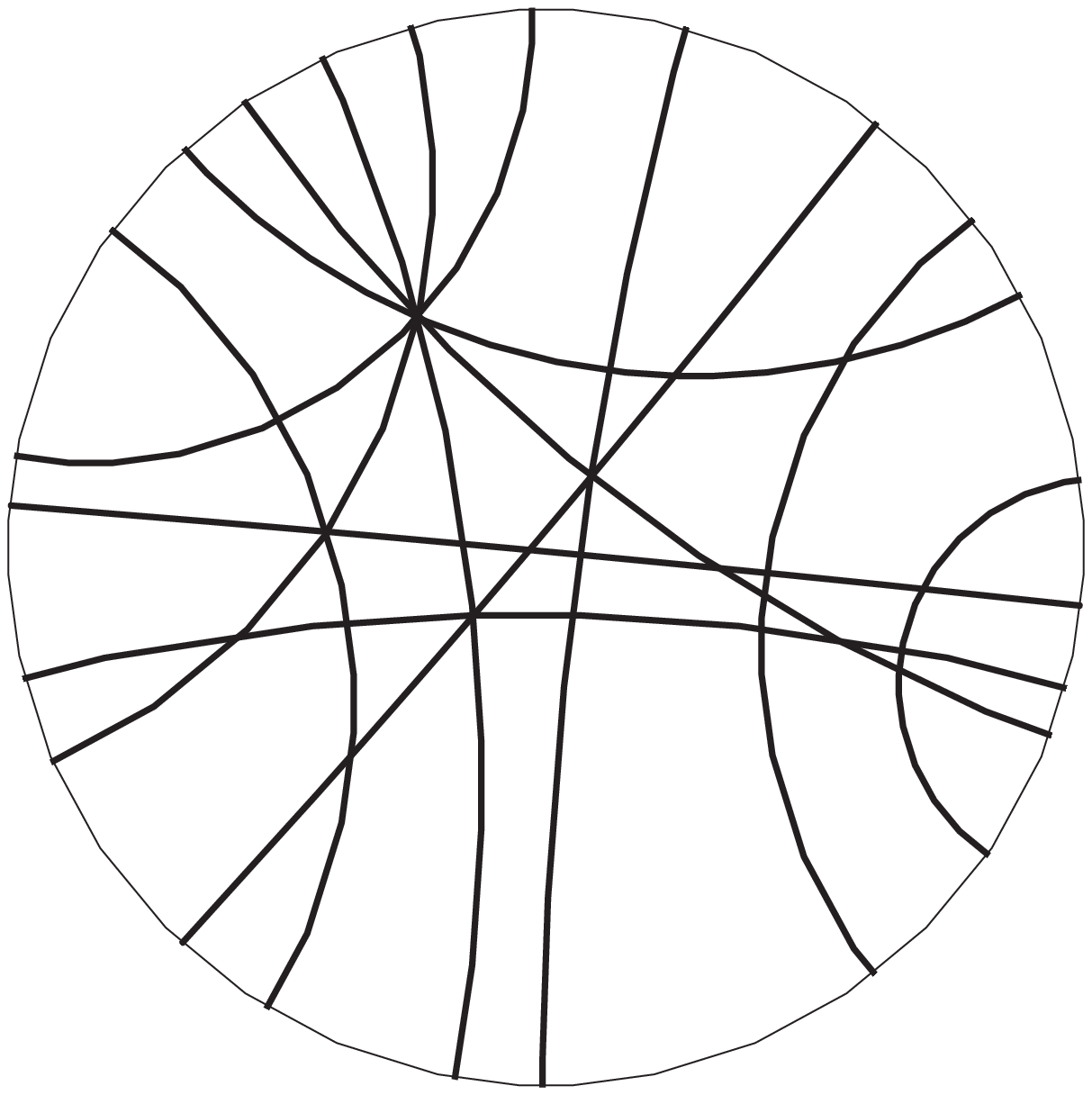}
\qquad
\includegraphics[width=2.5in]{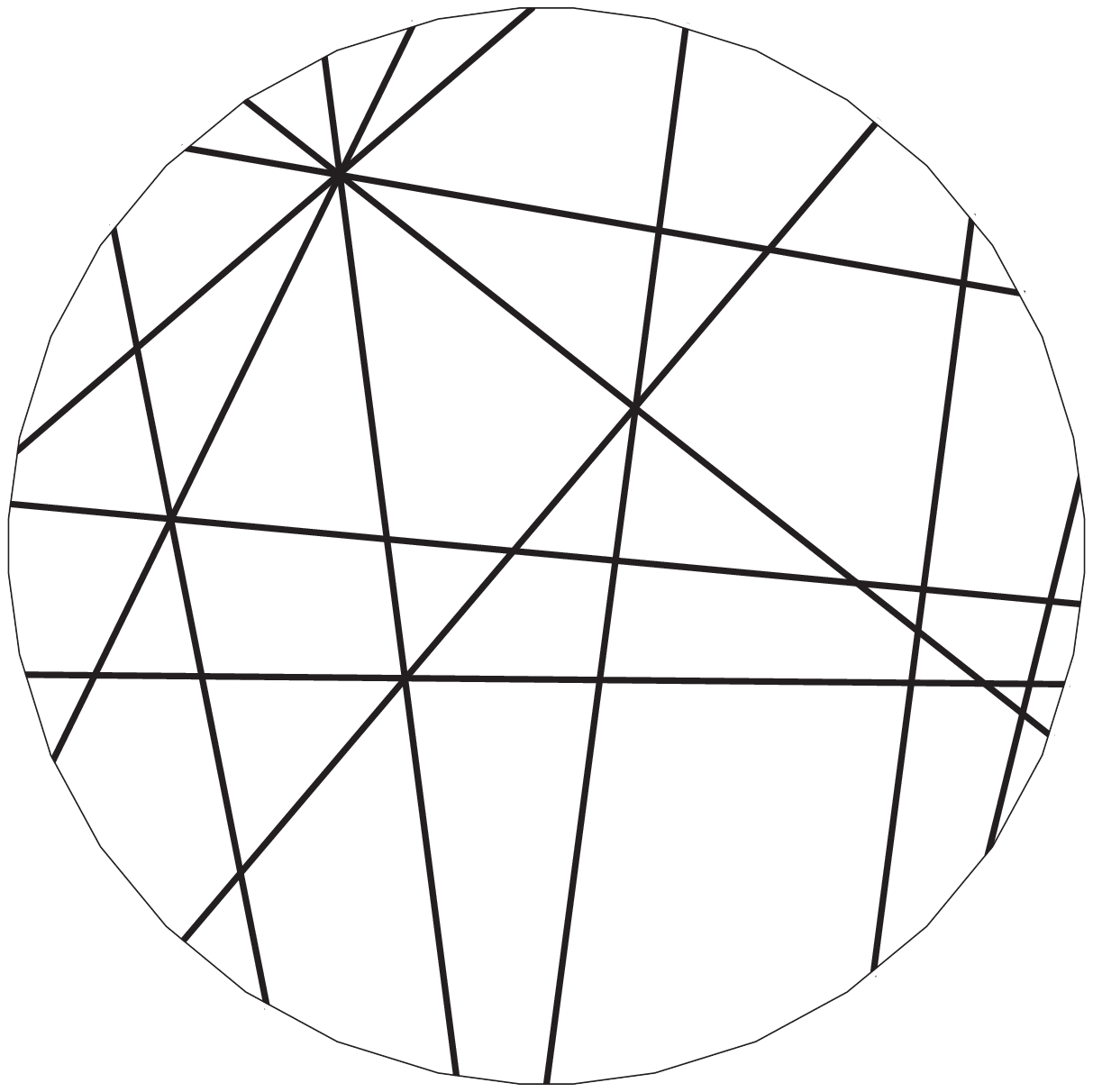}
\caption{Poincar\'e (left) and Klein (right) models of the hyperbolic
plane. Analogous models exist for any higher dimensional hyperbolic
space.}
\label{fig:poinklein}
\end{figure}

\begin{figure}[t]
\centering
\includegraphics[width=2.5in]{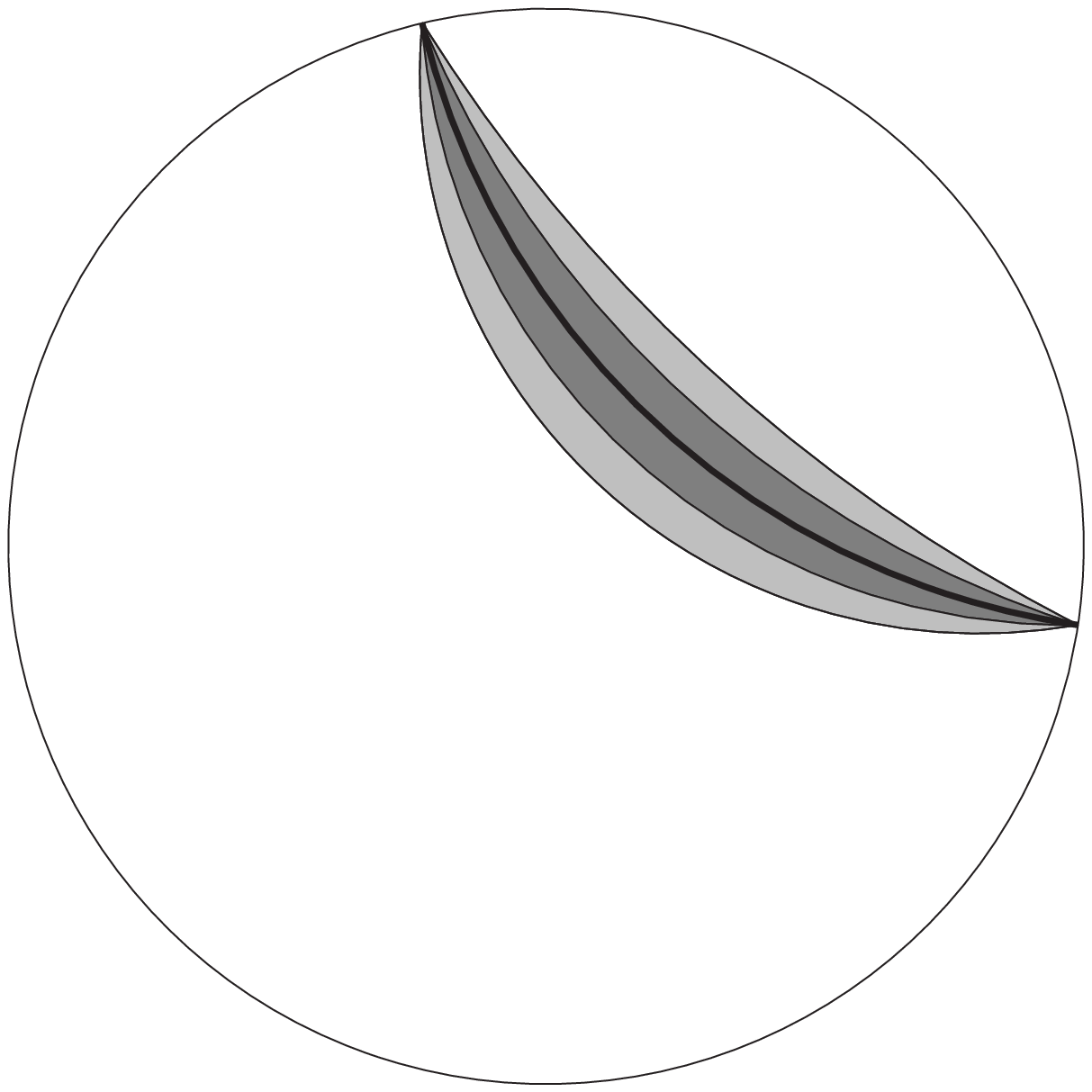}
\caption{Hyperspheres equidistant from a hyperbolic line, and
hyperbolically convex regions (shaded) of points within a given
distance bound from the line.}
\label{fig:bananas}
\end{figure}

Although our problem statements involve Euclidean and
spherical geometry, our solutions involve techniques from hyperbolic
geometry, and in particular the classical methods of embedding
hyperbolic space into Euclidean space: the {\em Poincar\'e model}, the
{\em halfspace model} and the {\em Klein model}.  For a general
introduction to these topics, see~\cite{Ive-92}.

In both the Poincar\'e and Klein models, the $d$-dimensional hyperbolic
space
$\H^d$ is viewed as homeomorphic to an open unit ball in a Euclidean space
$\E^d$, while the unit sphere bounding the ball forms a set of points
``at infinity''. In the Poincar\'e model (Figure~\ref{fig:poinklein},
left), the geodesics (lines) of the hyperbolic space are modeled by
circular arcs of the Euclidean space, perpendicular to the unit sphere. 
Hyperplanes are modeled by spheres perpendicular to the unit sphere, and
hyperbolic spheres are modeled by spheres fully contained within the unit
ball.  There are two more classes of surface that the Poincar\'e model
models as Euclidean spheres: {\em hyperspheres} (surfaces at constant
distance from a hyperplane) are modeled by spheres crossing the unit
sphere  non-perpendicularly, and {\em horospheres} are modeled by spheres
tangent to the unit sphere.  The Poincar\'e model thus preserves
spherical shape as well as the angles between pairs of curves or
surfaces. Any horosphere or hypersphere divides $\H^d$ into a convex and
a nonconvex region; we define a {\em horoball} or {\em hyperball} to be
the convex region bounded by a horosphere or hypersphere respectively.

An important variation of the Poincar\'e model is the
{\em halfspace model}, in which $\H^d$ is viewed as homeomorphic to a
Euclidean halfspace.  The unit sphere
at infinity in the Poincar\'e model corresponds to the boundary of the
halfspace, a (conventionally horizontal) hyperplane which is augmented by
an additional ``point at infinity''. As in the Poincar\'e
model, generic geodesics are modeled as semicircles
perpendicular to the halfspace boundary. Generic spheres, horospheres,
hyperplanes, and hyperspheres are modeled as Euclidean spheres
interior to the halfspace, tangent to its boundary, or crossing its
boundary perpendicularly or nonperpendicularly respectively. However,
geodesics containing the point at infinity are instead modeled by vertical
Euclidean lines, horospheres containing the point at infinity are modeled
by Euclidean hyperplanes parallel to the halfspace boundary, hyperplanes
containing the point at infinity are modeled by Euclidean hyperplanes
perpendicular to the halfspace boundary, and hyperspheres containing the
point at infinity are modeled by non-vertical Euclidean hyperplanes.

In the Klein model (Figure~\ref{fig:poinklein}, right), the geodesics of
the hyperbolic space map to line segments of the Euclidean space, formed
by intersecting Euclidean lines with the unit ball.  Hyperplanes thus map
to hyperplanes, and convex bodies map to convex bodies.  Although the
Klein model does not preserve spherical shape, it does preserve flatness
and convexity.  In the Klein model, spheres are modeled as ellipsoids
contained in the unit ball, horospheres are modeled as ellipsoids
tangent at one point to the unit ball, and hyperspheres are modeled as
halves of ellipsoids tangent in a $(d-1)$-sphere to the unit ball
(pairs of hyperspheres symmetric by reflection across a hyperplane match
up to form full ellipsoids in the Klein model).
Spheres, hyperspheres, and horospheres are not convex in $\H^d$
since they map to curved surfaces in the Klein model, but
balls, hyperballs, and horoballs are convex.

The Poincar\'e and Klein models of hyperbolic space are not intrinsic to
the space, rather there can be many such models for the same space.
The choice of model is determined by the selection of the hyperbolic point
mapped to the unit ball's center in the model, and by an orientation of
the space around this center.  We call this central point the {\em
viewpoint} since it determines the Euclidean view of the hyperbolic
space.

The connection between hyperbolic space and M\"obius transformations is
this: the isometries of the hyperbolic space are in one-to-one
correspondence with the subset of M\"obius transformations of the unit
ball that map the unit ball to itself, where the correspondence is
given by the Poincar\'e model~\cite[Theorem 6.3]{Ive-92}.  Any hyperbolic
isometry (or unit-ball preserving M\"obius transformation) can be
factored into a hyperbolic translation mapping some point of the
hyperbolic space to the viewpoint, followed by a rotation around the
viewpoint~\cite[Lemma 6.4]{Ive-92}.  Since rotation does not change the
Euclidean shape of the objects on the model, our problems of selecting an
optimal M\"obius transformation of a Euclidean or spherical space can all
be rephrased in terms of finding an optimal viewpoint of a hyperbolic
space.

\subsection{Hyperbolic Neighborhoods}

\begin{figure}[t]
\centering
\includegraphics[width=3.5in]{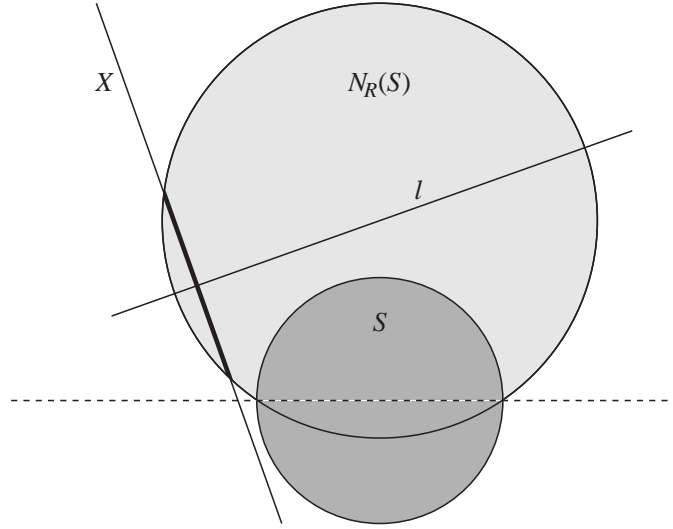}
\caption{Objects for proof of Lemma~\ref{lem:hyperconv}:
halfspace model of $\H^d$ (region above dashed line),
hypersphere $X$, halfspace $S$ (dark shaded area),
hyperball $N_R(S)$ (light shaded area),
Euclidean line $\ell$, and intersection
$X\cap N_R(S)$ (thick line segment).}
\label{fig:hyperconv}
\end{figure}

If $X$ is a hypersphere in $\H^d$, let $F(X)$ (the {\em flattening} of
$X$) denote the hyperplane that has the same points at infinity;
that is, as viewed in the Poincar\'e model, $X$ and $F(X)$ form spheres
that intersect the unit sphere in the same set of points.  Then there
exists a real number $R$ such that each point of $X$ is at distance $R$
from $F(X)$; we call $R$ the {\em hyperradius} of $X$.

For any real $R>0$, define the $R$-neighborhood $N_R(S)$ of a set $S$ to
be the set of points in $\H^d$ with distance at most $R$ from $S$.
We will mainly consider neighborhoods of hyperplanes and lines.

\begin{lemma}
If $S$ is convex, $N_R(S)$ is convex.
\end{lemma}

\begin{proof}
By the convexity-preserving properties of the Klein model, $S$ can be
represented as an intersection of halfspaces
$\bigcap_{x\in I} H_x$ for some index set $I$.
But then $N_R(S)$ is an intersection $\bigcap N_R(H_x)$ of
convex hyperballs.
\end{proof}

If $S$ is a hyperplane, $N_R(S)$ is the lens-shaped region
formed by the intersection of two hyperballs, with hyperradius $R$, each
of which has $S$ as the flattening of its boundary.
If $S$ is a line, $N_R(S)$ is the banana-shaped region formed by the
intersection of all hyperballs with hyperradius $R$ such that
the flattened boundary of the hyperball contains $S$.
Figure~\ref{fig:bananas} depicts neighborhoods of a line (which is also
a hyperplane) in $\H^2$.

For any horosphere $S$, one can find a halfspace model of $\H^d$ in which
$S$ is modeled as a plane parallel to the boundary of the halfspace; thus
$S$ naturally acquires a Euclidean geometry.  For a hypersphere $X$, on
the other hand, the natural intrinsic geometry is hyperbolic, and can be
found by mapping $X$ perpendicularly onto $F(X)$.
If we choose a halfspace model in which the infinite points of $X$ form
a $(d-2)$-flat in $\E^d$, then $X$ is mapped to a (non-vertical) Euclidean
hyperplane in the halfspace model, and the perpendicular projection is
modeled very simply as a Euclidean rotation around the $(d-2)$-flat into
the vertical hyperplane
$F(X)$.  For instance, with $d=3$, this halfspace model lets us view
perpendicular projection from $X$ to $F(X)$ in $\H^d$ as the rotation of
a non-vertical plane into a vertical plane around the line where the two
planes intersect the halfspace boundary.  Thus, a circular or convex set
in
$X$ projects into a circular or convex set in $F(X)$ respectively.  We
must be careful, however, with convexity in $X$: since $X$ is not itself
convex, it is possible for a convex set $K$ in $\H^d$ to intersect $X$ in
a set $K\cap X$ which is not convex, even when viewed according to the
intrinsic geometry within $X$. The next lemma describes a situation in
which this sort of pathological behavior can be shown not to happen.

\begin{lemma}
\label{lem:hyperconv}
Let $X$ be a hypersphere with hyperradius $r$,
let $S$ be a convex set,
and let $R>r$.
Then $N_R(S)\cap X$ is convex in the intrinsic geometry for $X$.
\end{lemma}

\begin{proof}
Since we can represent $S$ as an intersection of halfspaces, we can
represent $N_R(S)$ as an intersection of hyperradius-$R$ hyperballs.
By the convexity preserving property of intersection, the result
follows from its special case in which $S$ is a halfspace and $N_R(S)$
is a hyperball.  So, for the rest of the proof, we assume that we have
this special case.

Form a halfspace model of $\H^d$ by choosing as the point at infinity a
point $x$ that is on the boundary of $X$ and exterior to $N_R(S)$;
if no such point exists, then all of $X$ is contained in $N_R(S)$
so the intersection is trivially convex.
In this halfspace model, as illustrated in Figure~\ref{fig:hyperconv},
$X$ is modeled as a tilted Euclidean
hyperplane, while $N_R(S)$ is modeled as a Euclidean ball, so their
intersection is modeled by a Euclidean disk, which must
correspond to one of five types of sets in the intrinsic hyperbolic
geometry on $X$: a ball, horoball, hyperball, halfspace, or the
complement of a hyperball.  We must rule out the last of these cases,
which is not convex, and would be modeled by a Euclidean disk
that is less than half contained within the halfspace model.

In the halfspace model, hyperradius is
modeled by the angle at which a surface meets the boundary of the
model, so the condition that
$R>r$ implies that the hyperplane modeling $X$ meets the boundary more
steeply than does the ball modeling $N_R(S)$.
Consider the Euclidean line $l$ through the Euclidean center of the
ball modeling $N_R(S)$, and perpendicular to the Euclidean hyperplane
modeling
$X$. Because of this relation on the angles of the two surfaces, the two
points where $l$ meets the boundary of $N_R(S)$ are both contained
within the halfspace modeling $\H^d$.  Therefore, the Euclidean center
of the disk modeling $N_R(S)\cap X$ (which lies on $l$) is also contained
in the halfspace, so the portion of the disk contained in the halfspace is
more than half. Thus, the disk cannot model a hyperball complement, and
must be one of the other four possibilities, which are all convex.
\end{proof}

\subsection{Quasiconvex Programming}
\label{sec:qcp}

The viewpoint we seek in our optimal M\"obius transformation problems will
be expressed as the pointwise maximum of a finite set of
{\em quasiconvex functions}; that is, functions for which the level sets
are all convex.  To find this point, we use a generalized
linear programming framework of Amenta et al.~\cite{AmeBerEpp-Algs-99}
called {\em quasiconvex programming}.

A {\em generalized linear program} (or {\em LP-type
problem}) consists of a finite set
$S$ of {\em constraints} and an {\em objective function} $f$ mapping
subsets of $S$ to some totally ordered space and satisfying the
following properties:
\begin{enumerate}
\item For any $A\subset B$, $f(A)\le f(B)$.
\item For any $A$, $p$, and $q$,
if $f(A)=f(A\union\{p\})=f(A\union\{q\})$, then
$f(A)=f(A\union\{p,q\})$.
\end{enumerate}
The problem is to compute
$f(S)$ using only evaluations of $f$ on small subsets of~$S$.
A {\em basis} of a GLP is a set $B$
such that for any $A\subsetneq B$, $f(A)<f(B)$.
The {\em dimension} of a GLP is the maximum
cardinality of a basis.

Define a {\em nested convex family} to be a map
$\kappa(t)$ from the nonnegative real numbers to compact convex sets in
$\E^d$ such that if
$a<b$ then
$\kappa(a)\subset\kappa(b)$, and such that
for all $t$, $\kappa(t)=\bigcap_{t'>t}\kappa(t')$.
Any nested convex family $\kappa$ determines
a function $f_\kappa(x) = \inf\,\{\,t \mathrel{|} x \in \kappa(t)\,\}$
on $\R^d$, with level sets that are the boundaries of $\kappa(t)$.
If $f_\kappa$ does not take a constant value on any open set,
and if $\kappa(t')$ is contained in the interior of
$\kappa(t)$ for any $t'<t$, we say that $\kappa$ is {\em continuously
shrinking}.
Conversely, the level sets of any quasiconvex function form the
boundaries of the convex sets in a nested convex family, and if the
function is continuous and not constant on any open set then the family
will be continuously shrinking.

If $S=\{\kappa_1,\kappa_2,\ldots \kappa_n\}$ is a set of  nested convex
families, and $A\subset S$, let
$$
f(A)=\inf\Big\{\,(t,x) \mathrel{\big|}
x\in \mathop{\textstyle\bigcap}\limits_{\kappa_i
\in A}\kappa_i(t)
\Big\}
$$
where the infimum is taken in the lexicographic ordering,
first by $t$ and then by the coordinates of~$x$.
Amenta et al.~\cite{AmeBerEpp-Algs-99} define a {\em quasiconvex program}
to be a finite set $S$ of nested convex families,
with the objective function $f$ described above.

\begin{lemma}[Amenta et al.~\cite{AmeBerEpp-Algs-99}]
\label{lem:qcp}
Any quasiconvex program forms a generalized linear program of dimension at
most $2d+1$. If each $\kappa_i\in S$ is
constant or continuously shrinking, the dimension is at most $d+1$.
\end{lemma}

Due to the convexity-preserving properties of the Klein model,
we can replace $\E^d$ by $\H^d$ in the definition of a nested convex
family without changing the above result.

Generalized linear programs (and hence also quasiconvex programs) with
constant dimension can be solved by any of several
algorithms~\cite{AdlSha-MP-93,Ame-DCG-94,Cla-JACM-95,DyeFri-MP-89,Gae-SJC-95,MatShaWel-Algo-96},
the best of which use a linear number of calls to a subroutine for testing
whether a proposed solution remains valid after adding another
constraint, and a logarithmic number of calls to a subroutine for finding
a new basis of a constant-sized subproblem. 
We can also perform a more direct local optimization procedure for
quasiconvex programs: we can find $f(S)$ by applying steepest
descent, nested binary search, or other local optimization techniques
to find the point $x$ minimizing $f(x)=\max_i f_{\kappa_i}(x)$.

\section{Algorithms}
\label{sec:algs}

We now describe how to apply the quasiconvex programming framework
described above in order to solve our optimal M\"obius transformation
problems.  In each case, we form a set of nested convex families
$\kappa_i$, where each family corresponds to a function $f_i$ describing
the size of one of the objects in the problem (e.g., the transformed
radius of a sphere) as a function of the viewpoint location.  The
solution to the resulting quasiconvex program then gives the viewpoint
maximizing the minimum of these function values.  The only remaining
question in applying this technique is to show that, for each of the
problems we study, the functions of interest do indeed have convex level
sets.

\subsection{Maximizing the Minimum Sphere}
\label{sec:optsphere}

We begin with the simplest of the problems described in the
introduction: finding a M\"obius transformation that takes the unit ball
to itself and maximizes the minimum radius among a set of transformed
spheres.  Equivalently, we are given a set of spheres in a hyperbolic
space, and wish to choose a viewpoint for a Poincar\'e model of the
space that maximizes the minimum Euclidean radius of the spheres in the
model.

By symmetry, the radius of a sphere in the Poincar\'e model depends only
on its hyperbolic radius, and on the hyperbolic distance from its center
to the viewpoint.  Further, the modeled radius is monotonic in the
center-viewpoint distance.  Thus, if we let $f_i(z)$ denote the modeled
radius, as a function of the viewpoint location $z$, then the level sets
of $f_i$ form the boundaries of a nested convex family of concentric balls
in
$\H^d$.  As a limiting case, the level sets for the modeled radius of a
horosphere also form the boundaries of a nested convex family of
horoballs, all tangent to the same point at infinity.

\begin{theorem}
\label{thm:E-maxmin-radius}
Suppose we are given as input a set of $n$ spheres, all contained in the
unit ball in $\E^d$.  Then we can find the M\"obius
transformation of the unit ball that maximizes the minimum radius of the
transformed spheres, in $O(n)$ time, by solving a quasiconvex program
with one nested convex family per sphere.
\end{theorem}

Since all the nested convex families of this problem are continuously
shrinking, the resulting generalized linear program has dimension at
most $d+1$.

We note that the assumption that the spheres are contained in the unit
ball is necessary: the level sets for the modeled radius of a
hypersphere in $\H^d$ can be nonconvex, and in particular the locus of
viewpoints at which the hypersphere is modeled by a flat Euclidean plane
is itself a hypersphere.

If we view the unit sphere itself as being one of the input spheres,
then Theorem~\ref{thm:E-maxmin-radius} can be viewed as minimizing the
ratio between the radii of the largest and smallest transformed spheres.

\begin{open}
Is there an efficient algorithm for finding a M\"obius transformation
of $\E^d$ minimizing the ratio between the radii of the largest and
smallest transformed spheres, when the input does not necessarily include
one sphere that contains all the others?
\end{open}

We can also prove a similar result for radius optimization on the sphere:

\begin{theorem}
\label{thm:S-maxmin-radius}
Suppose we are given as input a set of $n$ spheres in $\S^d$.  Then we can
find the M\"obius transformation of $\S^d$ that maximizes the
minimum radius of the transformed spheres, in $O(n)$ time, by solving a
quasiconvex program with one nested convex family per sphere.
\end{theorem}

\begin{proof}
Embed $\S^d$ as the unit sphere in $\R^{d+1}$,
and consider it to be the sphere at infinity of a Poincar\'e
model of a hyperbolic space
$\H^{d+1}$. Then each of the input spheres $s_i$ is the boundary of a
unique hyperplane $h_i$ in $\H^{d+1}$.
For each input sphere, define a function $f_i(z)$ giving the (spherical
or Euclidean) radius of the image of the sphere under a M\"obius
transformation preserving the unit ball and taking $z$  to the viewpoint
of the transformed Poincar\'e model.
Then by symmetry, $f_i(z)$ depends only on the hyperbolic distance from
$z$ to $h_i$, so its level sets are the lens-shaped neighborhoods of
$h_i$ (Figure~\ref{fig:bananas}). Thus we again have a nested convex
family
$\kappa_i$ and can solve our optimal viewpoint problem as a quasiconvex
program.
\end{proof}

In this case, the dimension of the generalized linear program is at
most $d+2$.

\subsection{Spherical Vertex Separation}
\label{sec:optarc}

We next consider problems of using M\"obius transformations to separate
a collection of points.  We begin with the simpler version of this
problem, where the points are on a sphere.

\begin{theorem}
\label{thm:S-edge-length}
Suppose we are given as input a graph with $n$ vertices and $m$ edges,
and with each vertex assigned to a point on the sphere $\S^d$.  Then we
can find the M\"obius transformation of
$\S^d$ that maximizes the minimum arc length of the transformed graph
edges, in $O(m)$ time, by solving a quasiconvex program with one nested
convex family per edge.
\end{theorem}

\begin{proof}
Embed $\S^d$ as the unit sphere in $\R^{d+1}$,
and consider it to be the sphere at infinity of a Poincar\'e
model of a hyperbolic space
$\H^{d+1}$. Then the endpoints of each input edge $uv$
form the (infinite) endpoints of a unique hyperbolic line $\ell_{uv}$
in $\H^{d+1}$.
For each input edge, define a function $f_{uv}(z)$ giving the arc length
of the image of the edge under a M\"obius transformation preserving the
unit ball and taking $z$  to the viewpoint of the transformed Poincar\'e
model. Then by symmetry, $f_i(z)$ depends only on the hyperbolic distance
from
$z$ to $\ell_{uv}$, so its level sets are the
banana-shaped neighborhoods of
$\ell_{uv}$ (Figure~\ref{fig:bananas}). Thus we again have a nested
convex family $\kappa_i$ and can solve our optimal viewpoint problem as a
quasiconvex program.
\end{proof}

As in Theorem~\ref{thm:S-maxmin-radius}, the dimension of the
generalized linear program is $d+2$.

Similarly, we can find the
M\"obius transformation maximizing the minimum distance among a set of
$n$ transformed points by applying Theorem~\ref{thm:S-edge-length} to
the complete graph $K_n$.  However, the input size in this case is $n$,
while the algorithm of Theorem~\ref{thm:S-edge-length} takes time
proportional to the number of edges in $K_n$,
$O(n^2)$.  With care we can reduce the time to near-linear:

\begin{theorem}
Suppose we are given $n$ points in $\S^2$.
Then we can find the M\"obius transformation that maximizes the minimum
distance among the transformed points in $O(n\log n)$ time, by solving a
quasiconvex program with $O(n)$ constraints.
\end{theorem}

\begin{proof}
The Delaunay triangulation of the points can be computed in $O(n\log n)$
time, is M\"obius-invariant (due to its definition in terms of empty
circles), forms a planar graph with $O(n)$ edges, and is guaranteed to
contain the shortest transformed distance among the points.  Therefore,
applying Theorem~\ref{thm:S-edge-length} to the Delaunay triangulation
gives the desired result.
\end{proof}

In higher dimensions, the Delaunay triangulation may be complete, and so
gives us no advantage.  However we can again reduce the time from
quadratic by using a random sampling scheme similar to one from
our work on inverse parametric optimization problems~\cite{Epp-FOCS-99}:

\begin{theorem}
\label{thm:randomized-complete}
Suppose we are given $n$ points in $\S^d$.
Then we can find the M\"obius transformation that maximizes the minimum
distance among the transformed points in randomized expected time
$O(n\log n)$.
\end{theorem}

\begin{proof}
Initialize a graph $G$ to be empty, then repeat the following process:
choose a set $S$ of $O(n)$ random pairs of vertices, apply
Theorem~\ref{thm:S-edge-length} to maximize the minimum distance $\Delta$
among pairs in $G\cup S$, and then add to $G$ all pairs of transformed
vertices that have distance less than $\Delta$.

Each iteration of the process adds a set of edges to $G$ with expected
cardinality $O(n)$~\cite{Cla-JACM-95}, including at least one edge
involved in the optimal basis of the problem for the complete graph.
Therefore, the algorithm terminates in $O(1)$ iterations after having
solved $O(1)$ quasiconvex programs with expected size $O(n)$.

The pairs closer than $\Delta$ can be listed in time $O(n\log n +
k)$, where $k$ is the number of
pairs~\cite{BenStaWil-IPL-77,DicEpp-CGTA-96}.
\end{proof}

\begin{open}
Is there an efficient deterministic algorithm for maximizing the minimum
distance among $n$ points in $\S^d$, $d\ge 3$?
\end{open}

\subsection{Planar Vertex Separation}

As we now describe, we can find analogous results to the ones in the
previous section for a graph embedded in the unit ball in $\E^d$.  The
algorithms are essentially the same as in the previous section; the
difficulty in this case, however, is proving quasiconvexity.

\begin{theorem}
\label{thm:E-maxmin-dist}
Suppose we are given as input a graph with $n$ vertices and $m$ edges,
and with each vertex assigned to a point in the unit ball in $\E^d$.  Then
we can find the M\"obius transformation of
the unit ball that maximizes the minimum length of the transformed
graph edges, in $O(m)$ time, by solving a quasiconvex program with one
nested convex family per edge.
For a complete graph in $\E^2$, we can solve the problem in
deterministic time $O(n\log n)$,
and in $\E^d$, $d>2$, we can solve the problem in $O(n\log n)$
randomized expected time.
\end{theorem}

\begin{figure}[t]
\centering
\includegraphics[height=2in]{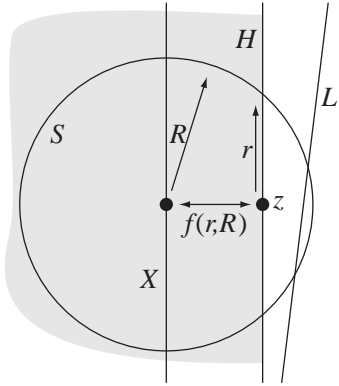}
\caption{An $(r,R)$-good point on $H$.}
\label{fig:good}
\end{figure}

\begin{proof}
We view the unit ball as a Poincar\'e model of $\H^d$, turning the
problem into one of finding the optimal viewpoint for a different
Poincar\'e model of the same hyperbolic space.  As we now show, there is
a convex set of viewpoints such that a particular edge $uv$ has some
length $\ell$ or greater, so the optimal viewpoint selection problem forms
a quasiconvex program.

To see this, we return to our unit ball in $\E^d$.  We view the whole of
$\E^d$ as the boundary of a halfspace model of $\H^{d+1}$.
There is a unique halfspace $H$ in $\H^{d+1}$ that intersects
$\E^d$ in the complement of our original unit ball.
Replace each edge $uv$ by a
hyperbolic line; in the halfspace model, this is a semicircle
perpendicular to $\E^d$, disjoint from $H$, and having endpoints at $u$
and $v$. Any M\"obius transformation of the unit ball can be extended
uniquely to an isometry of $\H^{d+1}$ that preserves $H$.
For each such transformation, consider the Euclidean hyperplane $E$ in the
halfspace model, parallel to the halfspace boundary and tangent to the
smallest semicircle; the (Euclidean) height of $E$ is just
half the length of the shortest transformed edge.  So another way of
describing our problem is that we are seeking the transformation for
which $E$ is as high as possible; that is, it intersects
the boundary of $H$ in a $(d-1)$-sphere of minimum (hyperbolic) radius.
Viewed as a hyperbolic object,
$E$ is just a horosphere.
Thus, we can describe our problem
in purely hyperbolic terms: we are given a halfspace
$H$, and a collection of hyperbolic lines. We seek a horosphere, with
its infinite point in
$H$, which intersects the boundary of $H$ in a minimum radius 
$(d-1)$-sphere and which touches all the lines.

In order to analyze the level sets for this problem, let two values $r<R$
be given, and from now on fix both halfspace $H$ and line $L$ in
$\H^{d+1}$.  We say that a point $z$ on the boundary of $H$ is
$(r,R)$-good if there exists a sphere $S$ in $\H^{d+1}$, with radius $R$,
with its center in $H$, with $S$ touching $L$, and
with $S\cap H$ forming a radius-$r$  $(d-1)$-sphere centered at $z$
(Figure~\ref{fig:good}). We wish to show that the set of
$(r,R)$-good points is convex.  Let $X$ be the hypersphere in $H$
containing the centers of radius-$R$ spheres that intersect the boundary
of $H$ in radius-$r$ $(d-1)$-spheres; then the hyperradius of $X$ is some
function $f(r,R)<R$. The set of $(r,R)$-good points is then just the
projection  onto the boundary of $H$ of $N_R(L)\cap X$.
So, the convexity of this set follows immediately from
Lemma~\ref{lem:hyperconv}.

Similarly, define
a $(r,\infty)$-good point to be the center of a radius-$r$
$(d-1)$-sphere on the boundary of $H$, such that some horosphere with
tangency in $H$ intersects the boundary at that $(d-1)$-sphere and
touches $L$. Then the set of $(r,\infty)$-good points is just the
intersection of the sets of $(r,R)$-good points, as $R$ ranges over all
positive real numbers. Thus, it is also convex.

Let $V$ be
the set of viewpoints in the original hyperbolic space $\H^d$
for which a given pair $uv$ maps to points at distance at least $\ell$.
Then $V$ is the perpendicular projection from the
boundary of $H$ of the set of
$(r,\infty)$-good points, where $r$ is some monotonic function of
$\ell$.  Since projection preserves convexity, this set is convex,
so the problem of selecting an optimal viewpoint forms a quasiconvex
program.
\end{proof}

\section{Applications}
\label{sec:apps}

\subsection{Spherical Graph Drawing}
\label{sec:gdraw}

\begin{figure}[t]
\centering
\includegraphics[height=2in]{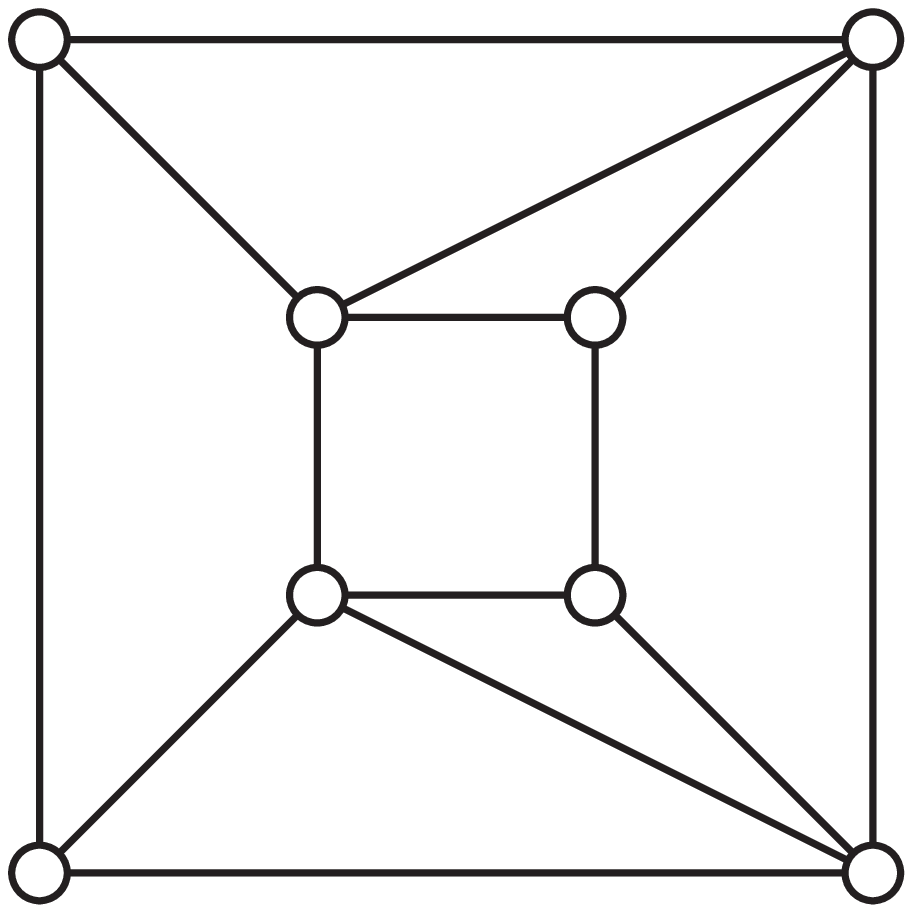}
\qquad
\includegraphics[height=2in]{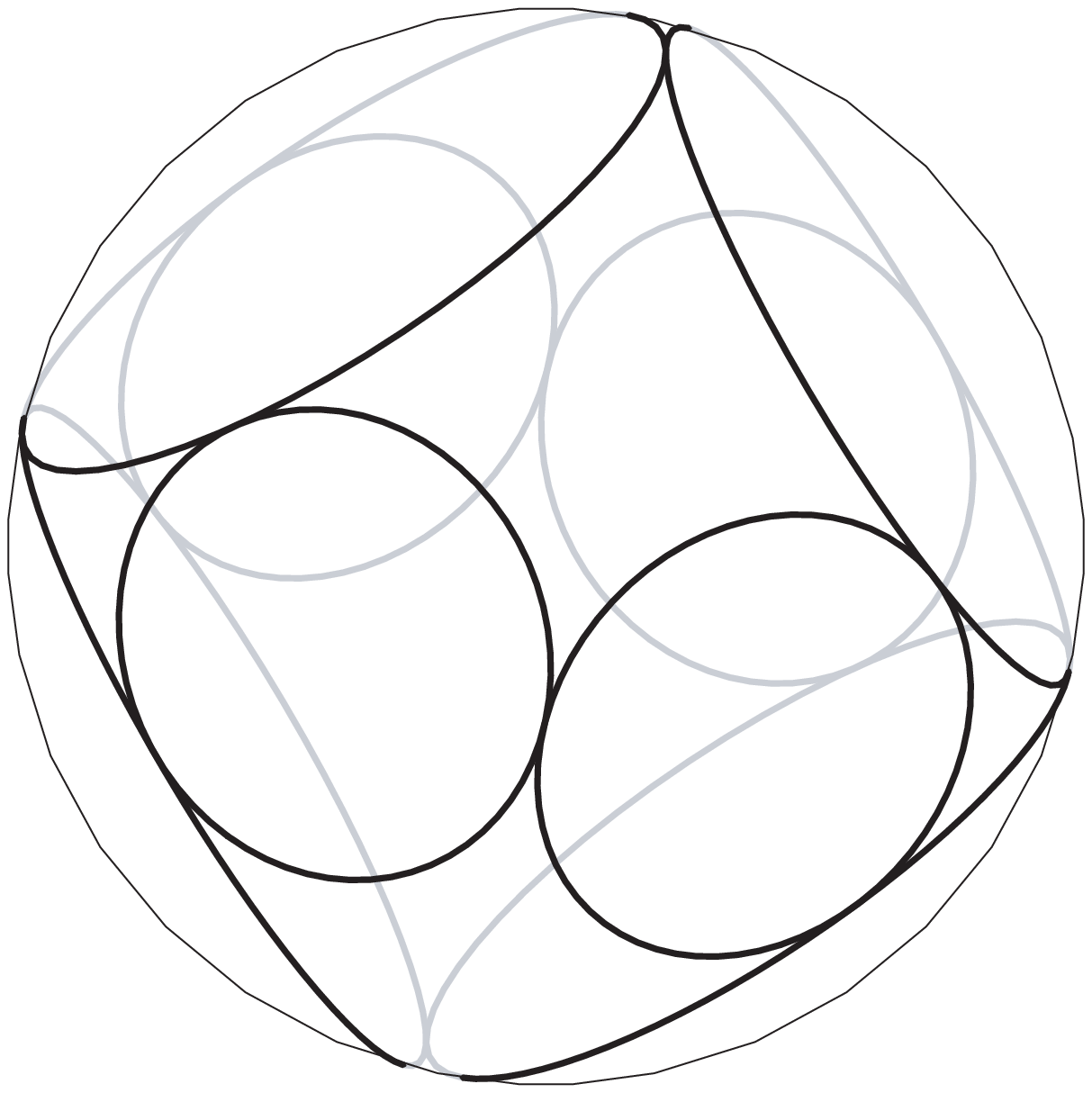}
\caption{A planar graph (left) and its coin graph
representation (right).}
\label{fig:coingraph}
\end{figure}

As is by now well known, any planar graph can be represented by a set of
disjoint circles in $\S^2$, such that two vertices are
adjacent exactly when the corresponding two circles are
tangent~\cite{BriSch-SJDM-93,Koe-BSAW-36,Sac-DM-94}.  We call such a
representation a {\em coin graph}; Figure~\ref{fig:coingraph} shows an
example.  Although it seems difficult to represent the positions of the
coins exactly, fast algorithms for computing numerical approximations to
their positions are known~\cite{ColSte-97,Moh-DM-93,Smi-91}.
By polar projection, we can transform any coin graph representation on
the sphere to one in the plane or vice versa.
See Ken Stephenson's web site
\url{http://www.math.utk.edu/~kens/} for more information, including
software for constructing coin graph representations and a bibliography
of circle packing papers.

It is natural to ask for the planar or spherical coin graph representation
in which all circles are most nearly the same size. However
it is NP-hard to determine whether a planar coin graph representation
exists in which all circles are equal, or in which the ratio between the
maximum and minimum radius satisfies a given
bound~\cite{BreKir-GD-95}.\footnote{The same result implies that it is
NP-hard to recognize coin graphs in three or more dimensions: $G$ is a
unit-disk coin graph if and only if $G\cup\{u,v\}$ is a
three-dimensional sphere tangency graph, where $u$ and $v$ are new
vertices adjacent to each other and to all vertices in $G$.}

However, if the graph is maximal planar, its coin graph representation
is unique up to M\"obius transformation, and we can apply
Theorem~\ref{thm:S-maxmin-radius} to find the optimal spherical coin graph
representation.  There is also a natural way of getting a canonical coin
graph representation from a non-maximal embedded planar graph:
add a new vertex in each face, connected to all the vertices of the
face.  Find the coin graph representation of the augmented graph, and
delete the circles representing the added vertices.  Again,
Theorem~\ref{thm:S-maxmin-radius} can then find the optimal M\"obius
transformation of the resulting coin graph.

Due to the fact that a
quasiconvex program only has a single global optimum, the transformed
coin graph will display any symmetries present in the initial graph
embedding.  That is, any homeomorphism of the sphere
that transforms the initial embedded graph into itself becomes simply a
rotation or reflection of the sphere in the optimal embedding. If the
graph has a unique embedding (i.e. is a subdivision of a 3-connected
planar graph) then any isomorphism of the graph becomes a rotation or
reflection.  For instance, the coin graph representation in
Figure~\ref{fig:coingraph} (right) has the full symmetry of the
underlying graph, while the planar drawing on the left of the figure
does not show the symmetries that switch the vertices in the inner and
outer squares.

Alternatively, by representing each vertex by the center of its circle,
a coin graph representation can be used to find a straight-line drawing
of a planar graph, or a drawing on the sphere in which the edges appear
as non-crossing great-circle arcs.  The algorithms in
Section~\ref{sec:optarc} can then be used to find a representation
maximizing the minimum vertex separation, among all M\"obius
transformations of the initial vertex positions.
For this variation of the problem, it may be appropriate to require that
the M\"obius transformation preserve the graph embedding:
for each pair $(u,v)$ and $(u,w)$ of edges adjacent around a face of the
embedding, we add a constraint that the chosen viewpoint belong to a
halfspace defined by the points $u$, $v$, and $w$.  This
constraint forces these three vertices to maintain their orientation
after the transformation is applied.  These additional $O(n)$
constant nested convex families do not increase
the asymptotic complexity of our algorithms.

\subsection{Hyperbolic Browser}
\label{sec:hbrowse}

There has been quite a bit of recent work in the information visualization
community on {\em hyperbolic browsers}, techniques for using hyperbolic
space to aid in the visualization of large graphs or graph-like
structures such as the world-wide web~\cite{MunBur-VRML-95}.  In these
techniques, a graph is arranged within a hyperbolic
plane~\cite{LamRaoPir-CHI-95} or three-dimensional hyperbolic
space~\cite{Mun-CGA-97}, which is then rendered using tools such as the
Klein model, Poincar\'e model, or by rendering the graph as it would be
seen by a viewer within the hyperbolic space.
The main advantage of hyperbolic browsers is that they provide a
``fish-eye'' view~\cite{ForKel-GD-95} that allows both the details of
the current focus of interest and the overall structure of the graph to be
viewed simultaneously.  In addition, the homogeneous and isotropic
geometry of hyperbolic space allows for natural and smooth
navigation from one view to another.

Although there are many interesting problems in graph layout for
hyperbolic spaces, we are interested in a simpler question: where should
one place one's initial focus, in order to make the overall graph
structure as clear as possible?  Previous work has handled this problem
by the simplistic approach of laying out the graph using a rooted
subtree such as a breadth-first or depth-first tree, and then placing
the focus at the root of the tree.  This approach will work well
if the tree is balanced, but if some parts of the tree end up
bushier than others then the corresponding parts of the graph
may be given a much more crowded initial view.  Instead, we hope to use
our techniques to find a viewpoint that shows the whole graph as clearly
as possible.

We assume we are given a graph, with vertices placed in a hyperbolic
plane.  As
in~\cite{LamRaoPir-CHI-95}, we further assume that each vertex has a
circular {\em display region} where information related to that
vertex is displayed; different nodes may have display regions of
different sizes.  It is then straightforward to apply
Theorem~\ref{thm:E-maxmin-radius} to these display regions; the result is
a focus placement for the Poincar\'e model of the graph in which we have
maximized the minimum size of any display region.
Although not expressable in terms of M\"obius transformations, we can
similarly find the Klein model maximizing the minimum diameter or
width of a transformed circle, since the level sets of these functions
are again simply concentric balls.

By applying Theorem~\ref{thm:E-maxmin-dist}
we can instead choose the focus for a Poincar\'e model that
maximizes the minimum distance between vertices, either among pairs from
the given graph or among all possible pairs.

\begin{open}
Is there an efficient algorithm for choosing a Klein model of a
hyperbolically embedded graph that maximizes the minimum
Euclidean distance between adjacent vertices?
\end{open}

Alternatively, we can apply similar methods to find a focus in
3-dimensional hyperbolic space that maximizes the minimum solid angle
subtended by any display region, since as in
Theorem~\ref{thm:E-maxmin-radius} these angles are again quasiconvex
functions with spherical level sets.

\begin{open}
Does there exist an efficient algorithm to find a viewpoint in
3-dimensional hyperbolic space maximizing the minimum angle separating
any pair among
$n$ given points?
\end{open}

Even the 2-dimensional Euclidean version of this maxmin angle separation
problem is interesting~\cite{McKay-89}: the level sets are nonconvex
unions of two disks, so our quasiconvex programming techniques do not
seem to apply.

\subsection{Conformal Mesh Generation}
\label{sec:cmesh}

One of the standard methods of two-dimensional structured mesh
generation~\cite{BerPla-HCG-00,ThoWarMas-85} is based on {\em conformal
mapping} (that is, an angle-preserving homeomorphism).
The idea is to find a conformal map from the domain to
be meshed into some simpler shape such as a disk, use some predefined
template to form a mesh on the disk, and invert the map to lift the mesh
back to the original domain.  Conformal meshes have significant
advantages: the orthogonality of the grid lines means that one can avoid
certain additional terms in the definition of the partial differential
equation to be solved~\cite{ThoWarMas-85}.  Nevertheless, despite
much work on algorithms for finding conformal
maps~\cite{DriVav-SISC-98,How-PhD-90,Smi-91,SteSch-CMFT-97,Tre-SSC-80}
conformal methods are often avoided in favor of quasi-conformal mesh
generation techniques that allow some distortion of angles, but provide
greater control of node placement~\cite{BerPla-HCG-00,ThoWarMas-85}.

M\"obius transformations are conformal,
and any two conformal maps from a simply connected domain to a disk can
be related to each other via a M\"obius transformation.
However, different conformal maps will lead to different structured
meshes: the points of the domain mapped on or near the center of the disk
will generally be included in mesh elements with the finest level of
detail, and points near the boundary will be in coarser mesh elements.
Therefore, as we now describe, we can use our optimal M\"obius
transformation algorithms to find the conformal mesh that best fits the
desired level of detail at different parts of the domain, reducing the
number of mesh elements created and providing some of the node placement
control needed to use conformal meshing effectively.

We formalize the problem by assuming an input domain in which certain
interior points $p_i$ are marked with a desired element size $s_i$.
If we find a conformal map $f$ from the domain to a disk,
the gradiant of $f$ maps the marked element sizes
to desired sizes $s'_i$ in the transformed disk:
$s'_i = || f' (p_i) ||$.
We can then choose a structured mesh with element size $\min s'_i$
in the disk, and transform it back to a mesh of the original domain.
The goal is to choose our conformal map in a way that maximizes $\min
s'_i$, so that we can use a structured mesh with as few elements as
possible.  Another way of interpreting this is that
$s'_i$ can be seen as the radius of a small disk
at $f(p_i)$.  What we want is the viewpoint that maximizes the minimum
of these radii.

By applying a single conformal map, found using one of the
aforementioned techniques, we can assume without loss of generality that
the input domain is itself a disk.  Since the conformal maps from disks
to disks are just the M\"obius transformations, our task is then to find
the M\"obius transformation maximizing $\min s'_i$.  
Since $s'_i$ depends only on the hyperbolic distance of the viewpoint
from $p_i$, the level sets for this problem are themselves disks,
so we can solve this problem by the same quasiconvex programming
techniques as before.  Indeed, we can view this problem as a limiting
case of Theorem~\ref{thm:E-maxmin-radius} for infinitesimally small (but
unequal) sphere radii.

\begin{open}
Because of our use of a conformal map to a low aspect ratio shape (the
unit disk), rotation around the viewpoint does not significantly affect
element size.  Howell~\cite{How-PhD-90} describes methods for computing
conformal maps to high aspect ratio shapes such as rectangles.
Can one efficiently compute the optimal choice of conformal map to a
high-aspect-ratio rectangle to maximize the minimum desired element size?
What if the rectangle aspect ratio can also be chosen by the
optimization algorithm?
\end{open}

\subsection{Brain Flat Mapping}

In order to visualize and understand the complicated strucure of the
brain, neuroanatomists have sought methods for stretching its convoluted
surface folds onto a flat plane.  Hurdal et al.~\cite{HurBowSte-TR-99}
have proposed a principled way of performing this stretching, via
conformal maps: since the surfaces of major brain components such as
the cerebellum are topologically disks, the Riemann mapping theorem
proves the existence of a conformal map from these surfaces onto a
Euclidean unit disk, sphere, or hyperbolic plane.  Hurdal et al.{}
approximate this conformal map by using a fine triangular mesh to
represent the brain surface, and forming a coin graph representation of
this mesh.  Each triangle from the brain surface can then be mapped to
the triangle connecting the corresponding three coin centers.
For details, and further references to brain flat mapping techniques,
see their paper.

Necessarily, any flat mapping of a curved surface such as the brain's
involves some distortions of area, but the distortions produced by
conformal mapping can be severe; thus, it would be of interest to choose
the mapping in such a way that the distortion is minimized.  As we
already noted in section~\ref{sec:cmesh}, the remaining degrees of
freedom in choosing a conformal mapping can be described by a
single M\"obius transformation.  Thus, we need to formulate a measure of
distortion, and find the transformation optimizing that measure.

Since we want to measure area, and the mapping constructed by the method
of Hurdal et al. is performed on triangles of a mesh, the most natural
quality measure for this purpose seems to be in terms of those
triangles: we want to minimize the maximum ratio $a/a'$ where $a$ is the
area of a triangle in the initial three-dimensional map, and $a'$ is the
area of its image in the flat map.  Unfortunately, we have not yet been
able to extend our techniques to this quality measure.  A positive
answer to the following question would allow us to apply our quasiconvex
programming algorithms:

\begin{open}
Let $T$ be a triangle in the unit disk or sphere, and let $C$ be the set
of viewpoints for M\"obius transformations that transform $T$ into a
triangle of area at least $A$.  Is $C$ necessarily convex?
\end{open}

Instead of attempting to optimize the area of the triangles, it seems
simpler (although perhaps less accurate) to optimize the area of the disks
in the coin graph.  Under the assumption that the initial triangular mesh
has elements of roughly uniform size, it would be desirable that the coin
graph representation similarly uses disks of as uniform size as possible.
This can be achieved by our linear-time algorithms by applying
Theorem~\ref{thm:E-maxmin-radius} in the unit disk, or
Theorem~\ref{thm:S-maxmin-radius} in the sphere.
In case the triangular mesh is nonuniform, it may be appropriate to
apply a weighted version of these theorems, where the weight of each
disk is computed from the lengths of edges incident to the corresponding
mesh vertex.

\section{Conclusions}

We have identified several applications in information visualization
and structured mesh generation for which it is of interest to find a
M\"obius transformation that optimizes an objective function, typically
defined as the minimum size of a collection of geometric objects.
Further, we have shown that these problems can be solved either by
local optimization techniques, or by linear-time quasiconvex programming
algorithms.  For the problems where the input to the quasiconvex program
is itself superlinear in size (maximizing the minimum distance between
transformed points) we have described Delaunay triangulation and random
sampling techniques for solving the problems in near-linear time.

Several theoretical open problems arising in our investigations have
been enumerated throughout the paper.  There is also an important
problem in the practical application of our algorithms:
although there should be little difficulty implementing local
optimization techniques for our problems, the linear-time quasiconvex
programming algorithms are based on two primitives that
(while constant time by general principles) have not been specified in
sufficient detail for an implementation, one to test a new constraint
against a given basis and the other to find the changed basis of a set
formed by adding a new constraint to a basis.  If the basis
representation includes the value of the objective function, testing a
new constraint is simply a matter of evaluating the corresponding object
size and comparing it to the previous value.  However, the less frequent
basis change operations require a more detailed examination of the
detailed structure of each problem, which we have not carried out.  For
an example of the difficulty of this step for a different problem
(the minimum area ellipse containing points in $\E^2$),
see~\cite{GaeSch-IPL-98}.  In practice it may be appropriate to
combine the two approaches, using local optimization techniques to find
a numerical approximation to the basis change operations needed for the
quasiconvex programming algorithms.  Especially in the coin graph
application, the input to the quasiconvex program is already itself a
numerical approximation, so this further level of approximation should not
cause additional problems, but one would need to verify that a
quasiconvex programming algorithm can behave robustly with approximate
primitives.

More generally, we are unaware of previous
computational geometry algorithms involving hyperbolic geometry,
although many Euclidean constructions (such as the Delaunay
triangulation or hyperplane arrangements) can be translated
to the hyperbolic case without difficulty using the Poincar\'e or Klein
models.  We expect many other interesting problems and
algorithms to be discovered in this area.

\small\raggedright
\bibliographystyle{abuser}
\bibliography{mobius}

\end{document}